\documentclass[prb,twocolumn,showpacs,preprintnumbers,amsmath,amssymb]{revtex4}
\usepackage{graphicx} 
\usepackage{color}
\usepackage{dcolumn} 
\usepackage{bm} 

\newcommand{\be}{\begin{equation}}
\newcommand{\ee}{\end{equation}}
\newcommand{\bea}{\begin{eqnarray}}
\newcommand{\eea}{\end{eqnarray}}

\newcommand{\reff}{\text{eff}}

\newcommand{\s}{\sigma}

\newcommand{\up}{\uparrow}
\newcommand{\down}{\downarrow}
\newcommand{\la}{\langle}
\newcommand{\ra}{\rangle}

\newcommand{\bml}{\begin{mathletters}}
\newcommand{\eml}{\end{mathletters} \hspace{-5pt}}

\begin{document}
\title{Band-Insulator-Metal-Mott-Insulator transition in the\\
half--filled $t-t^{\prime}$
ionic-Hubbard chain}

\author{G.I. Japaridze$^{1,2}$, R. Hayn$^{3}$, P. Lombardo$^{3}$ and E. M\"uller-Hartmann$^{1}$}
\affiliation{$^{1}$ Institut f\"ur Theoretische Physik,
Universit\"at zu K\"oln,
D-50937 K\"oln, Germany\\
$^{2}$Andronikashvili Institute of Physics, Tamarashvili 6, 0177
Tbilisi, Georgia \\
$^{3}$Laboratoire Mat\'eriaux et Micro\'electronique de Provence
associ\'e au Centre National de la Recherche Scientifique.
UMR 6137. Universit\'e de Provence, France}

\date{\today}

\begin{abstract}

We investigate the ground state phase diagram of the half-filled
$t-t^{\prime}$ repulsive Hubbard model in the presence of a
staggered ionic potential $\Delta$, using the continuum-limit
bosonization approach. We find, that with increasing
on-site-repulsion $U$, depending on the value of the
next-nearest-hopping amplitude $t^{\prime}$, the model shows three
different versions of the ground state phase diagram. For
$t^{\prime} < t^{\prime}_{\ast}$, the ground state phase diagram
consists of the following three insulating phases: Band-Insulator at
$U<U_{c}$, Ferroelectric Insulator at $U_{c} < U < U_{s}$ and
correlated (Mott) Insulator at $U > U_{c}$. For $t^{\prime} >
t^{\prime}_{c}$ there is only one transition from a spin gapped
metallic phase at $U<U_{c}$ to a ferroelectric insulator at $U >
U_{c}$. Finally, for intermediate values of the next-nearest-hopping
amplitude $t^{\prime}_{\ast} < t^{\prime} < t^{\prime}_{c}$ we find
that with increasing on-site repulsion, at $U_{c1}$ the model
undergoes a second-order commensurate-incommensurate type
transition from a band insulator into a metallic state and at larger
$U_{c2}$ there is a Kosterlitz-Thouless type transition from a metal
into a ferroelectric insulator.

\end{abstract}

\pacs{71.10.Fd, 71.27.+a, 71.30.+h}

\maketitle

\section{Introduction}

During the last decades, the Mott metal-insulator transition has
been a subject of great
interest.\cite{Mott_Book_90,Gebhard_Book_97, IFT_98} In the
canonical model for this transition -- the single-band Hubbard model
--  the origin of the insulating behavior is the on-site Coulomb
repulsion between electrons. For an average density of one electron
per site, the transition from the metallic to the insulating phase
is expected to occur {\it with increasing on-site repulsion} when
the electron-electron interaction strength $U$ exceeds a critical
value $U_{c}$, which is usually of the order of the
delocalization energy. Although the underlying mechanism driving the
Mott transition is by now well understood, many questions remain
open, especially about the region close to the transition point
where perturbative approaches fail to provide reliable answers.

The situation is more fortunate in one dimension, where
non-perturbative analytical methods together with well-controlled
numerical approaches allow to obtain an almost complete description
of the Hubbard model and its dynamical properties.\cite{EFGKK_2005}
However, even in one dimension, apart from the exactly solvable
cases, a full treatment of the fundamental issues related to the
Mott transition still constitutes a hard and challenging problem.

Intensive recent activity is focused on studies of the extended
versions of the Hubbard model which display, with increasing Coulomb
repulsion, a transition from a band-insulator (BI) into the
correlated (Mott) insulator phase. Various models considered include
those, which show a continuous evolution from a BI into the MI
phase\cite{Rosch_06,Anfuzo_Rosch_06,Monien_06} as well as those, where the
transformation of a BI to a correlated (Mott) insulator takes place
via a sequence of quantum phase
transitions.\cite{Fab_99,Brune_03,Man_04,Otsu_05,TAJN_06,Randeria_06,LHJ_06,Dagotto_06}

Intensive recent activity is focused on studies of the extended
versions of the Hubbard model with alternating on-site energies $\pm
\Delta$, known as the ionic Hubbard model (IHM).
\cite{Fab_99,Brune_03,Man_04,Otsu_05,TAJN_06,Randeria_06,LHJ_06,Dagotto_06,Tor_01,Aligia_04,Aligia_Hallberg_05,Batista_Aligia_05,Scalettar_06}
The model has a long-term history,\cite{Hub_Torr_81} however the
increased current interest widely comes from the possibility to
describe the interaction driven BI to MI transition within one
model. In one dimension this evolution with increasing on-site
repulsion is characterized by two quantum phase transitions: first a
(charge) transition from the BI to a ferroelectric insulator (FI)
and second, with further increased repulsion, a (spin) transition
from the FI to a correlated MI.\cite{Fab_99} Detailed numerical
studies of the 1D IHM clearly show, that an unconventional metallic
phase is realized in the ground state of the model only at the
charge transition point and the BI and MI phases are separated in
the phase diagram by the insulating ferroelectric
phase.\cite{Brune_03,Man_04,Otsu_05,TAJN_06} Studies of the 2D IHM
using the cluster dynamical mean field theory, reveal a similar
phase diagram.\cite{Dagotto_06}

On the other hand, recent studies of the IHM using the dynamical
mean field theory (DMFT) approach, show that in high dimensions the
BI phase can be separated from the MI phase by the finite stripe of
a metallic phase.\cite{Randeria_06} Moreover, recent studies of the
IHM with site diagonal disorder using the DMFT approach, also show
the existence of a metallic phase which separates the BI phase from
the MI phase in the ground state phase diagram of the disordered
IHM.\cite{LHJ_06} It looks so, that in low-dimensional models with
perfect nesting of the Fermi surface, the metallic phase is reduced
to the charge transition line, while in higher dimensions the space
for realization of a metallic phase opens.  Note that the very
presence of a metallic phase along critical lines separating two
insulating phases is common for 1D systems with competing
short-range interactions responsible for the dynamical generation of
a charge gap.\cite{Emery_79} However, one-dimensional models of
correlated electrons, showing with increasing electron-electron
coupling a transition from an insulating to a metallic phase are
less known.\cite{Note_1}

In this paper we show, that the one-dimensional half-filled
ionic-Hubbard model supplemented with the next-nearest-neighbor
hopping term ($t^{\prime}$) is possibly the simplest one-dimensional model
of correlated-electrons, which shows a ferroelectric ground state in a wide
area of the phase diagram easily controlled by the model parameters.

We also show that in a certain range of the model parameters
the $t-t^{\prime}$ ionic-Hubbard chain shows, with increasing Coulomb repulsion,
a transition from a band-insulator to a metal and, with further increase of the Hubbard repulsion, a transition from a metal to a ferroelectric insulator (see Fig.1).
\begin{figure}[hb]
\vspace{0.2cm}
\includegraphics[width =7.0cm,angle=0]{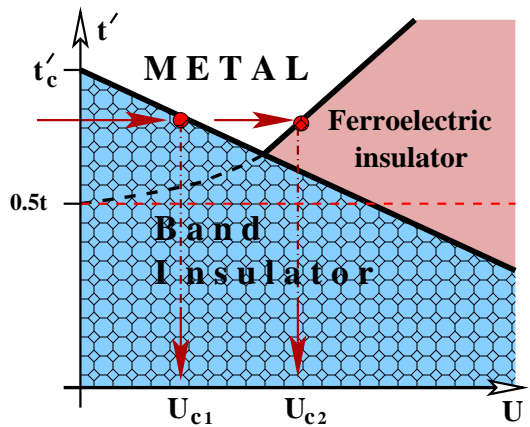}
\caption{Qualitative phase diagram of the half-filled $t-t^{\prime}$ ionic-Hubbard chain for
$t^{\prime}>0.5t$ and weak and moderate values of the on-site Hubbard repulsion $U$. The parameter
$t^{\prime}_{c}=0.5t \sqrt{1 + (\Delta/4t)^{2}}+\Delta/8$ corresponds to the insulator to metal transition point
in the free ionic chain, where $\Delta/2$ is the amplitude of alternating ionic potential. The dashed curve
line marks the I-M transition line in the $t-t^{\prime}$ Hubbard model, corresponding to the case $\Delta=0$. The dashed line
$t^{\prime}=0.5t$ is given as an eye guide.
\label{Fig:Fig1}}
\end{figure}

The paper is organized as follows. In Section II, the model and its
several important limiting cases are briefly reviewed. In the
Section III the weak-coupling bosonization description is obtained.
The results are summarized in Section IV.


\section{\bf The model}

The Hamiltonian we consider is given by
\begin{eqnarray}\label{t1t2_IH_model}
{\cal H} &=& -t \,\sum_{n,\sigma} \left (c^{\dagger}_{n,
\sigma}c^{\phantom{\dagger}}_{n+1,\sigma} \, + \,
c^{\dagger}_{n+1,\sigma}c^{\phantom{\dagger}}_{n, \sigma}\right
)\nonumber \\
& + & t^{\prime} \, \sum_{n,\sigma} \left
(c^{\dagger}_{n,\sigma}c^{\phantom{\dagger}}_{n+2,\sigma} +
c^{\dagger}_{n+2,\sigma}c^{\phantom{\dagger}}_{n, \sigma}\right )\nonumber \\
&&\hspace{-8mm}+ \sum_{n,\sigma}\left( \delta\mu +
(-1)^{n}\frac{\Delta}{2}\right)\rho_{n,\sigma}\,+ U \, \sum_{n} \,
\rho_{n,\up}\rho_{n,\down}\, .
\end{eqnarray}
Here $c^{\dagger}_{n, \sigma}$ $(c^{\phantom{\dagger}}_{n,\sigma})$
are electron creation (annihilation) operators on site $n$ and, with
spin projection $\sigma =\up,\down$, $\rho_{n,\sigma}=
c^{\dagger}_{n,\sigma}c^{\phantom{\dagger}}_{n,\sigma}$. The
nearest-neighbor hopping amplitude is denoted by $t$, the
next-nearest-neighbor hopping amplitude by $t^{\prime}$
($t,t^{\prime}>0$), $\Delta$ is the potential energy difference
between neighboring sites, $U$ is the on-site Coulomb repulsion and
the band-filling is controlled by the proper shift of the chemical
potential $\delta\mu$. For $t^{\prime}=0$, we recover the
Hamiltonian of the ordinary ionic-Hubbard chain, while for
$\Delta=0$ the Hamiltonian of the $t-t^{\prime}$ Hubbard chain.

At $U=0$ the model is easily
diagonalized in momentum space to give a dispersion relation in the first Brillouin zone $-\pi/2 < k < \pi/2$
\begin{equation}\label{DR}
E_{\pm}(k) = 2t^{\prime}\cos 2k - \delta\mu \pm
\sqrt{4t^{2}\cos^{2}k + (\Delta/2)^{2}} \, .
\end{equation}

Let us first analyze the dispersion relation (\ref{DR}). For
$t,t^{\prime}>0$ the absolute maximum of the upper band is reached
at $k=0$, while the absolute minimum
\be\label{E+MIN}
 E^{+}_{min}= -2t^{\prime}- \delta\mu + \Delta/2\, ,
\ee
at $k=\pm \pi/2$.

The lower band shows a more complicated dependence on the model
parameters. For
\be\label{t^prime_ast} t^{\prime} \,< \,  t^{\prime}_{\ast} \, = \,0.5t
~\sqrt{ 1 + (\Delta/4t)^{2}}-\Delta/8
\ee
the absolute minimum of the lower band is reached at $k=0$ and while
the absolute maximum at $k=\pm \pi/2$ and is equal to
\be \label{E-Max-1} E^{-}_{max}= -2t^{\prime} - \delta\mu -
\Delta/2\, . \ee
In the case of half-filling the lower band is completely filled
and the upper band is empty. The system is a band-insulator with a
gap in the excitation spectrum
\be \label{Emin1} \Delta_{exc} \equiv E^{+}_{min}-
E^{-}_{max}=\Delta . \ee
The corresponding shift of the chemical potential $\delta\mu =
-2t^{\prime} $ is easily determined from the condition $E^{+}_{min}
+ E^{-}_{max}=0$. Thus, for $t^{\prime}<t^{\prime}_{\ast}$ the
ground state and low-energy excitation spectrum of the model are not
affected by the increase of the next-nearest hopping amplitude
$t^{\prime}$.

The effect of $t^{\prime}$ becomes nontrivial for $t^{\prime}\,
> \, t^{\prime}_{\ast}$, when the absolute maximum of the lower band
$$
E^{-}_{max}=2t^{\prime} - \delta\mu - \sqrt{4t^{2} + (\Delta/2)^{2}}
\,
$$
is reached at $k=0$. For
\be\label{tc-IM} t^{\prime} \,< \, t^{\prime}_{c}\equiv \,
0.5t~\sqrt{1 + (\Delta/4t)^{2}} + \Delta/8 \ee
the absolute maximum of the lower band at $k=0$ remains lower than
the absolute minimum of the upper band at $k=\pi/2$. Therefore the
system remains an insulator, but the indirect gap
\be \label{ExcitaGap2} \Delta_{B} = \frac{\Delta}{2}\, + \,
\sqrt{4t^{2} + (\Delta/2)^{2}} - 4t^{\prime} \ee
decays linearly with increasing $t^{\prime}$ and finally vanishes at
$t^{\prime}=t^{\prime}_{c}$. It is straightforward to find, that the
corresponding shift of the chemical potential which ensures
half-filling in this case is given by $\delta\mu =
-2t^{\prime}_{\ast}$.

At $t^{\prime} = t^{\prime}_{c}$ the $t-t^{\prime}$ ionic chain
experiences a transition from a band-insulator to a metal. For
$t^{\prime} > t^{\prime}_{c}$ a gapless phase is realized,
corresponding to a metallic state with four Fermi points $\pm
k_{F1}$ and $\pm k_{F2}$, which satisfy the relation
$2(k_{F1}-k_{F2})=\pi$.

At $U  > 0$ several limiting cases of the model
(\ref{t1t2_IH_model}) have been the subject of intensive current
studies. In particular intensive recent activity has been focused on
studies of the ground state phase diagram of the ionic-Hubbard model
(IHM), corresponding to the limiting case $t^{\prime}=0$ and of the
$t-t^{\prime}$ Hubbard model corresponding to the limiting case
$\Delta=0$.

Current interest in this model mostly originates from the
possibility to describe the evolution from the band insulator (BI)
at $U \ll \Delta$ into a correlated (Mott) insulator (MI) for $U \gg
\Delta$ within a single system.
\cite{Fab_99,Tor_01,Brune_03,Man_04,Aligia_04,Aligia_Hallberg_05,Batista_Aligia_05,Otsu_05,TAJN_06,Randeria_06,LHJ_06,Dagotto_06}
In the case of the one-dimensional ionic-Hubbard model this
evolution is characterized by  two quantum phase transitions in the
ground state. \cite{Fab_99} With increasing $U$ the first is a
charge transition at $U = U_{c}$, from the BI to a bond-ordered,
spontaneously dimerized, ferroelectric insulator (FI). At the
transition point the charge sector of the model becomes gapless,
however for $U>U_c$ the charge gap opens again. When $U$ is further
increased, the second (spin) transition from the FI phase into the
MI phase takes place at $U_{s}
> U_{c}$. At this transition the spin gap vanishes and the spin
sector remains gapless in the MI phase for $U > U_{s}$. A similar
ground state phase diagram has been recently established also in the
case of 2D IHM using the cluster dynamical mean field
theory.\cite{Dagotto_06} Thus, for low-dimensional versions of the
IHM with perfect nesting property of the Fermi surface, the
transition from a BI to a FI is characterized by the
presence of a metallic (charge gapless) phase only at the transition
line. In the ground state phase diagram of the BI phase is separated
from the MI phase by a FI phase.\cite{Fab_99,Man_04,TAJN_06,Dagotto_06}

The homogeneous half-filled $t-t^{\prime}$ Hubbard chain is a prototype model to
study the metal-insulator transition in one dimension and therefore
has been the subject of intensive studies in recent years.
\cite{Fabrizio_96,Kuroki_97,Fabrizio_98,DaulNoack_98,DaulNoack_00,AebBaerNoack_01,Gros_01,Gros_02,Torio_03,Gros_04,Fabrizio_04,JNB_06}
As in the case of the ionic-chain, in the $t-t^{\prime}$ Hubbard model
an increase of the next-nearest hopping $t^{\prime}$ changes the topology of
the Fermi surface: at half-filling and
for $t^{\prime}< 0.5t$, the electron band of has two Fermi points at
$ k_{F}=\pm \pi/2$, separated from each other by the umklapp vector
$q=\pi$. In this case, a weak-coupling renormalization group
analysis predicts the same behavior as for $t^{\prime}=0$ - the
dynamical generation of a charge gap for $U>0$, and gapless
magnetic excitations.\cite{Fabrizio_96}

For $t^{\prime}> 0.5t$, the Fermi level intersects the one-electron band at four points $\left(k_{F}^{\pm} \neq \pm \pi/2 \right)$. For weak Hubbard coupling ($U
\ll t$) the infrared behavior is governed by the low-energy
excitations in the vicinity of the four Fermi points, in full
analogy with the two-leg Hubbard model. \cite{BalentsFisher_96} The
Fermi vectors $k_{F}^{\pm}$ are sufficiently far from $\pi/2$ to
suppress first-order umklapp processes and therefore the charge
excitations are gapless, however the spin degrees of freedom are
becoming gapped.\cite{BalentsFisher_96,Fabrizio_96,DaulNoack_00,Gros_01,Torio_03}
Since at half-filling $4(k_{F}^{+}-k_{F}^{-})=2\pi$, with increasing
on-site repulsion higher-order umklapp processes become relevant
for intermediate values of $U$. Therefore, starting from a metallic
region for small $U$ at a given value of $t^{\prime}$ ($t^{\prime}>
0.5 t$), one reaches a transition line $U=U_{c}(t^{\prime})$ above
which the system is insulating with both charge and spin gaps.
\cite{Fabrizio_96}

As we will show in this paper for $t^{\prime} < t^{\prime}_{\ast}$,
where the topology of Fermi surface is restricted to two Fermi
points, the ground state phase diagram of the $t-t^{\prime}$
ionic-Hubbard chain coincides with that of the IHM, while for
$t^{\prime} > t^{\prime}_{c}$, where the model is characterized by
four Fermi points - it coincides with that of the  $t-t^{\prime}$
Hubbard chain. Most interesting is the case $t^{\prime}_{\ast} <
t^{\prime} < t^{\prime}_{c}$, where with increasing on-site
repulsion we observe two transitions in the ground state: at
$U_{c1}$ the model undergoes a second-order
commensurate-incommensurate type transition from a band insulator
into a metallic state and at large $U_{c2}$ there is a
Kosterlitz-Thouless type transition from a metal into a correlated
ferroelectric insulator.


\section{Bosonization results}

In this section we analyze the low-energy properties of the
$t-t^{\prime}$ ionic-Hubbard chain using the continuum-limit
bosonization approach. We first consider the regime $U, \Delta,
t^{\prime} \ll t$, linearize the spectrum in the vicinity of the
two Fermi points $k_{F}=\pm \pi/2$ and go to the continuum limit by
substituting
\be
c_{n\sigma} \rightarrow  \sqrt{a_{0}}\, \Big[ i^{n} \psi_{R\sigma}(x) +
(-i)^n \psi_{L\sigma}(x)\,\Big]\, , \label{linearization}
\ee
where $x=na_{0}$, $a_{0}$ is the lattice spacing, and
$\psi_{R\s}(x)$ and $\psi_{L\s}(x)$ describe right-moving and
left-moving particles, respectively. The chosen type of decoupling of the
model into "free" and "interaction" parts allows to treat the gap
"creating" ($\Delta$ and $U$) and gap "destructing"
($t^{\prime}$) terms on equal footing and reveals easily their
competition within the continuum-limit treatment.

The right and left fermionic fields are bosonized in the standard
way:\cite{GNT}
\bea \psi_{R \s}(x)&=&\frac{1}{\sqrt{2\pi a_{0}}} e^{{\it
i}\sqrt{\pi}[\Phi_{\s}(x)+\Theta_{\s}(x)]}
,\\
\psi_{L \s}(x)&=& \frac{1}{\sqrt{2\pi a_{0}}} e^{-{\it
i}\sqrt{\pi}[\Phi_{\s}(x)-\Theta_{\s}(x)]}\, , \eea
where $\Phi_{\s}(x)$ and $\Theta_{\s}(x)$ are dual bosonic fields,
$\partial_t \Phi_{\s} = v_{F} \partial_x \Theta_{\s}$ and
$v_{F}=2ta_{0}$.

This gives the following bosonized Hamiltonian:
$$ {\cal H}  =  {\cal H}_{\up} + {\cal H}_{\down}
+ {\cal
H}_{\up\down}\, ,
$$
where
\bea
{\cal H}_{\s} &=&  \int  dx \Big\{\frac{v_{F}}{2}\big[
\left(\partial_{x}\Phi_{\s}\right)^{2} +
\left(\partial_{x}\Theta_{\s}\right)^{2}\big]\nonumber\\
&&\hspace{-10mm} - \frac{\mu_{eff}} {\sqrt{\pi}}\partial_{x}\Phi_{\s}
- \frac{\Delta}{2\pi a_{0}}\sin\sqrt{4\pi}\Phi_{\s}\, \Big\}\,  \quad \left(\s =\up , \down \right)\label{H-sigma_BOS}
\eea
and
\bea
{\cal H}_{\up\down} &=&  \int  dx \Big\{\,\big[\, \frac{U}{\pi}
\partial_{x}\Phi_{\up} \partial_{x}\Phi_{\down}\nonumber\\
&& + \frac{U}{\pi^{2} a_{0}^{2}} \sin\sqrt{4\pi}\Phi_{\up}
\sin\sqrt{4\pi}\Phi_{\down}\,\big] \Big\}. \label{H-up-down_BOS}
\eea
Here we have introduced
\bea
& \mu_{{\it eff}} = 2t^{\prime} + \delta\mu  &\nonumber\\
& = \left\{
\begin{array}{l}
   \hspace*{0.6cm} 0 \hspace*{1.3cm} \mbox{for} \hspace*{0.5cm}
    t^{\prime} < t^{\prime}_{\ast} \\
    2\left(t^{\prime}-t^{\prime}_{\ast}\right)
    \hspace*{0.6cm} \mbox{for} \quad \hspace*{0.1cm}
    t^{\prime}_{\ast} < t^{\prime} <  t^{\prime}_{c} \\
\end{array}
\right . \, . & \label{ChemPot_gg} \eea

The next step is to introduce the charge
\begin{equation}
\varphi_{c} = {\textstyle \frac{1}{\sqrt{2}}} (\phi_{\uparrow} +
\phi_{\downarrow}), \qquad \vartheta_{c} = {\textstyle
\frac{1}{\sqrt{2}}} (\theta_{\uparrow} - \theta_{\downarrow })
\label{bos_carge}
\end{equation}
and spin fields
\begin{equation}
\varphi_{s} = {\textstyle \frac{1}{\sqrt{2}}} (\varphi_{\uparrow} -
\varphi_{\downarrow}),\qquad \vartheta_{s} = {\textstyle
\frac{1}{\sqrt{2}}} (\theta_{\uparrow} - \theta_{\downarrow})
\label{bos_spin}
\end{equation}
to describe corresponding degrees of freedom. After a simple
rescaling, we arrive at the bosonized version of the Hamiltonian
(\ref{t1t2_IH_model})
$$
{\cal H} = {\cal H}_{s} + {\cal H}_{c} + H_{cs}\, ,
$$
where
\bea
{\cal H}_{c} & = &  \int dx \Big\{\frac{v_{c}}{2K_{c}}(\partial_{x}\varphi_{c})^2
+\frac{v_{c}K_{c}}{2} (\partial_x \vartheta_{c})^2 \nonumber \\
&& - \mu_{\reff}\sqrt{\frac{2}{\pi}}\partial_{x}\varphi_{c}
- \frac{U}{2 \pi^{2} a_{0}^{2}}\cos(\sqrt{8\pi}\varphi_{c}) \Big\},
\label{SGc}\\
{\cal H}_{s} &= &  \int dx \Big\{\frac{v_{s}}{2}\big[(\partial_{x}\varphi_{s})^2
+\frac{1}{2}(\partial_x \vartheta_{s})^2\big]\nonumber \\
&&+ \frac{U}{2 \pi^{2} a_{0}^{2}}\cos(\sqrt{8\pi}\varphi_{s}) \Big\},
\label{SGs}\\
{\cal H}_{cs} & = &  - \frac{\Delta}{\pi a_{0}} \int dx \, \sin
\left( \sqrt{2\pi} \varphi_{c} \right) \cos \left( \sqrt{2\pi}
\varphi_{s}\right) \label{H_CS_BOS} \eea
with the charge stiffness parameter $K_{c} \simeq 1-U/4 \pi t $ at
$U/4\pi t \ll 1$.

\subsection{Non-interacting case}

To assess the accuracy of the continuum-limit treatment it is
instructive to start our bosonization analysis from the exactly
solvable case of the ionic-chain.

At $U=0$ the system is decoupled into the "up" and "down" spin
component parts ${\cal H} = {\cal H}_{\up} + {\cal H}_{\down}$,
where for each spin component the Hamiltonian is the sine-Gordon
model with topological term (\ref{H-sigma_BOS}). Each of these
Hamiltonians is the standard Hamiltonian for the
commensurate-incommensurate transition, which has been intensively
studied in the past using bosonization \cite{C_IC_transition} and
the Bethe ansatz.\cite{JNW_1984} This allows to apply the theory of
commensurate-incommensurate transitions to the metal-insulator
transition in the considered case of a half-filled $t-t^{\prime}$
chain with ionic distortion.

At $\mu_{eff}=0$, the model is described by the theory of two
commuting sine-Gordon fields ($\sin\beta\Phi_{\s}$) with
$\beta^{2}=4\pi$. In this case the excitation spectrum is gapped and
the excitation gap is given by the mass of the "up" ("down") field
soliton $M_{\up}=M_{\down}=\Delta/2$. In the ground state the
$\Phi_{\up}$ and $\Phi_{\down}$ fields are pinned with vacuum
expectation values $\langle 0|\Phi_{\s}|0 \rangle = \sqrt{\pi}(n +
1/4)$. Using the standard bosonized expression for the $2k_{F}$
modulated part of the charge density \cite{GNT}
\bea \rho_{c}(x) &\simeq&  (-1)^{n}\frac{1}{\pi
a_{0}}\sum_{\s}\sin(\sqrt{4\pi}\Phi_{\s}(x))\nonumber\\
&=& \frac{(-1)^{n}}{2\pi
a_{0}}\sin(\sqrt{2\pi}\varphi_{c}(x))\cos(\sqrt{2\pi}\varphi_{s}(x))
 \label{upDown_density} \eea
we obtain that at $\mu_{eff}=0$ the ground state of the system
corresponds to a CDW type band-insulator with a single energy scale
given by the ionic potential $\Delta$.

At $\mu_{eff} \neq 0$ it is necessary to consider the ground state
of the sine-Gordon model in sectors with nonzero topological charge.
The competition between the chemical potential term ($t^{\prime}$)
and the commensurability energy given by $\Delta$ finally drives a
continuous phase transition from a gapped (insulating) phase at
$\mu_{{\it eff}} < \mu_{{\it eff}}^{c}$ to a gapless (metallic)
phase at
\be\label{CriticalLine} \mu_{\reff} > \mu_{\reff}^{c} = \Delta/2\, .
\ee
Using (\ref{ChemPot_gg}) we easily obtain, that the critical value
of the n-n-n hopping amplitude $t^{\prime}$, obtained from the condition
(\ref{CriticalLine}) coincides with the exact value for the
ionic-chain given in (\ref{tc-IM}).

As we observe, the insulator-metal transition at $t^{\prime}>t^{\prime}_{c}$ is connected with a change of the topology of
the Fermi surface and a corresponding redistribution of the electrons
from the lower ("-") band into the upper ("+") band. We use as an order
parameter of this transition the number $N_{+}$ of electrons transferred
into the "+" band, which is related to the value of the new Fermi
point $k_{0} \sim \sqrt{\mu -\mu_{c}}$. At the
transition point the compressibility of the system is
\be
\partial E_{0}/\partial \mu \sim -k_{0}^{-1} = -(\mu -\mu_{c})^{-1/2}\,
,
\ee
showing an inverse square-root singularity, where $E_{0}$ is the ground state energy.

Before we start to consider the interacting case, it is useful to
continue our analysis of the noninteracting case, but within the basis of the charge and spin Bose fields, which is more convenient for interacting electrons.

The Hamiltonian we have to consider now is given by
\bea {\cal H} & = &  \int dx
\Big\{\frac{v_{F}}{2}\big[(\partial_{x}\varphi_{c})^2
+ (\partial_x \vartheta_{c})^2\big] - \mu_{\reff}\sqrt{\frac{2}{\pi}}\partial_{x}\varphi_{c} \nonumber \\
&&\hspace{10mm}+ \frac{v_{F}}{2}\big[(\partial_{x}\varphi_{s})^2
+\frac{1}{2}(\partial_x \vartheta_{s})^2 \big] \nonumber \\
&& \hspace{4mm}-\frac{\Delta}{\pi a_{0}} \sin \left( \sqrt{2\pi} \varphi_{c}
\right) \cos \left( \sqrt{2\pi} \varphi_{s}\right) \Big\}\,
.\label{H_CS_BOS_1} \eea

We decouple the interaction term in a mean-field manner by
introducing

\begin{eqnarray}
\label{m_{c}}
m_{c}&=& \Delta \cdot \langle \cos\sqrt{2\pi}\varphi_{s}\rangle\,\, ,\\
\label{m_{s}} m_{s}&=& \Delta\cdot \langle
\sin\sqrt{2\pi}\varphi_{c}\rangle\, ,
\end{eqnarray}
and get the mean-field bosonized version of the Hamiltonian ${\cal H}=
{\cal H}_{c}+{\cal H}_{s}$ which is
given by the two commuting quantum sine-Gordon models
\begin{eqnarray}
{\cal H}_{c}&=&\int dx\Big\{{v_{F} \over 2}[(\partial_{x}\vartheta_{c})^2 +(\partial_x
\varphi_{c})^2]\nonumber\\
&&- \mu_{\reff}\sqrt{\frac{2}{\pi}}\partial_{x}\varphi_{c} -\frac{m_{c}}{\pi a_{0}}\sin(\sqrt{2\pi} \varphi_{c})\Big\},
\label{Free_Bos_MF_Charge}\\
{\cal H}_{s}&=&\int dx\Big\{{v_{F} \over 2}[(\partial_{x}\vartheta_{s})^2 + (\partial_x
\varphi_{s})^2]\nonumber\\
&&-\frac{m_{s}}{\pi a_{0}}\cos(\sqrt{2\pi} \varphi_{s})\Big\}\, .
\label{Free_Bos_MF_Spin}
\end{eqnarray}

Although the mean-field Hamiltonian is once again given by the sum
of two decoupled sine-Gordon models (see Eq. (\ref{H-sigma_BOS})),
the dimensionality of the $\cos(\beta\phi)$ operators at
$\beta^{2}=2\pi$ and $\beta^{2}=4\pi$ are different. Therefore, in
marked contrast with the bosonized theory in terms of "up" and
"down" fields (\ref{H-sigma_BOS}), the pair of Hamiltonians given by
(\ref{Free_Bos_MF_Charge})-(\ref{Free_Bos_MF_Spin}) represents a
very complex basis to describe the BI phase i.e. the CDW state with
equal charge and spin gaps $\Delta_{c}=\Delta_{s}=\Delta/2$.

Nevertheless, below we use the benefit of the exact solution of
the sine-Gordon model to get a qualitatively and almost quantitatively
accurate description of the problem even in the "spin-charge" basis.
To see this, let us start from the case $\mu_{\reff}=0$. We will use
the following exact relations between the bare mass $m$ and the
soliton physical mass $M$ for the sine-Gordon theory with
$\beta^{2}=2\pi$\cite{Al_B_Zamolodchikov_95}
\be\label{Al-B-Z_1}
M/\Lambda = {\cal C}_{0}\left(m/\Lambda\right)^{2/3}\, ,
\ee
and the exact expression for the expectation value of the
$\cos\beta\phi$ field\cite{Luk_Zam_97}
\be\label{A-B-Z_1} \la \, \cos \sqrt{2\pi} \varphi \, \ra = {\cal
C}_{1}\left(M/\Lambda\right)^{1/2}\, . \ee
Here
\be \label{Al-B-Z_2}
{\cal C}_{0}=
\frac{2 \Gamma\left(1/6
\right)}{\sqrt{\pi}\Gamma\left(2/3\right)}
\left[\frac{\Gamma(3/4)}{2\Gamma(1/4)}\right]^{\frac{2}{3}}
\ee
and
\be\label{A-B-Z_2} {\cal C}_{1}=
\frac{2}{3}\left(\frac{3\pi}{4}\right)^{1/4}
\frac{\Gamma(3/4)}{\Gamma(1/4)}\, \ee
and $\Lambda=2t$ is the bandwidth.

Using (\ref{Al-B-Z_1})-(\ref{A-B-Z_2}) one easily finds that
\bea\label{SC_Mc}
 \Delta_{c}/\Lambda  &=& {\cal
C}_{0}\left(\Delta/\Lambda\right)^{2/3}\la \, \cos\sqrt{2\pi}
\varphi_{s} \, \ra^{2/3}\nonumber\\
& = & {\cal C}_{0}\,{\cal
C}_{1}^{2/3}\left(\Delta/\Lambda\right)^{2/3}\left(M_{s}/\Lambda\right)^{1/3}\,
\\
\label{SC_Ms} \Delta_{s}/\Lambda  &=&  {\cal
C}_{0}\left(\Delta/\Lambda\right)^{2/3}\la \, \sin\sqrt{2\pi}
\varphi_{c} \, \ra^{2/3}\nonumber\\
& = & {\cal C}_{0}\,{\cal
C}_{1}^{2/3}\left(\Delta/\Lambda\right)^{2/3}\left(M_{c}/\Lambda\right)^{1/3}
\, . \eea
The self-consistent solution of the Eqs. (\ref{SC_Mc})-(\ref{SC_Ms})
gives
\be\label{Delta_c-Delta_s}
\Delta_{c}=\Delta_{s} = \gamma \Delta/2
\ee
with $\gamma = 2{\cal C}^{3/2}_{0}{\cal C}_{1}= 0.94256...$. Thus
the mean-field treatment of the ionic-chain within the spin-charge
basis gives not only a qualitatively correct but, rather quantitatively
accurate description for the system in the BI phase.

For completeness of our description, let us now consider the
insulator to metal transition in the ionic-chain in the charge-spin
Bose field basis. The corresponding mean-field decoupled charge
Hamiltonian is given by the $\beta^2=2\pi$ quantum sine-Gordon model
with the topological term (\ref{Free_Bos_MF_Charge}). From the exact
solution of the SG model\cite{DHN} it is known that the excitation
spectrum of the model at $\beta ^{2}=2\pi$  consists of solitons and
antisolitons with mass $M_{c}$, and soliton-antisoliton bound states
("breathers") with masses $ M_{c}^{n=1}=2M_{c}\sin(\pi /6)=M_{c} $
and $ M_{c}^{n=2}=2M_{c}\sin(\pi /3)=\sqrt{3}M_{c} $. The transition
from the BI into the metallic phase takes place when the effective
chemical potential exceeds the mass of the lowest breather i.e. at
$\mu_{eff} = M_{c}=\Delta/2$. For $\mu_{eff}>M_{c}$ the vacuum
average of the charge field is not pinned at all, such that the spin
gap $\la M_{s} \ra \sim \langle
\sin(\sqrt{2\pi}\varphi_{c})\rangle=0$.

Thus, using the bosonization treatment we easily and with good
numerical accuracy describe the exactly solvable case of the
band-insulator to metal transition, which takes place in the
$t-t^{\prime}$ ionic chain with increasing
next-nearest-hopping amplitude $t^{\prime}$.

\subsection{Interacting case}

At $U \neq 0$ the use a similar mean-field decoupling allows us
to rewrite the Hamiltonian as two commuting double sine-Gordon
models $ {\cal H} = {\cal H}_{c} + {\cal H}_{s}$ where
\bea {\cal H}_{c} & = &  \int dx
\Big\{\frac{v_{c}}{2}\big[(\partial_{x}\varphi_{c})^2
+ (\partial_x \vartheta_{c})^2 \big] \nonumber \\
&& - \mu_{\reff}\sqrt{\frac{2K_{c}}{\pi}}\partial_{x}\varphi_{c} -
\frac{m^{r}_{c}}{\pi a_{0}}
\sin(\sqrt{2\pi K_{c}}\varphi_{c})\nonumber \\
&&-\frac{M_{c}}{2 \pi^{2} a_{0}^{2}}\cos(2\sqrt{2\pi
K_{c}}\varphi_{c}) \Big\},
\label{MF_DSG_Charge}\\
{\cal H}_{s} &= &  \int dx
\Big\{\big[\frac{v_{s}}{2}(\partial_{x}\varphi_{s})^2
 + (\partial_x \vartheta_{s})^2 \big]\nonumber \\
&&\hspace{-10mm}-\frac{m^{r}_{s}}{\pi a_{0}}\cos(\sqrt{2\pi} \varphi_{s}) +
\frac{M_{s}}{2 \pi^{2} a_{0}^{2}}\cos(\sqrt{8\pi}\varphi_{s})
\Big\}. \label{MF_DSG_Spin}  \eea
Here
\begin{eqnarray}
\label{m^R_{c}}
m^{r}_{c}&=&  \Delta_{r} \cdot \langle \cos\sqrt{2\pi}\varphi_{s}\rangle\,\, ,\\
\label{m^R_{s}} m^{r}_{s}&=& \Delta_{r}\cdot \langle
\sin\sqrt{2\pi}\varphi_{c}\rangle\, ,
\end{eqnarray}
$M_{c}$ and $M_{s}$ are the effective model parameters. The
renormalized band gap
\be\label{IRgaps1} \Delta_{r} = \Delta ~[1- (U/U^{\ast})\, ],
\ee
includes the Hartree renormalization of the band-gap by the on-site
repulsion $U$, where $U^{\ast}(\Delta)$ is a phenomenological parameter. For given $\Delta$, $U^{\ast}$ is of the order of $U_{c}$ in the standard IHM with the same
amplitude of the ionic distortion. The charge stiffness
parameter is $K_{c} <1$ in the case of repulsive interaction.

As we observe at $U > 0$ the charge sector is described by the
double frequency sine-Gordon model with strongly relevant basic and
marginally relevant double-field operators complemented by the
topological term. The spin sector is also given by the double
frequency sine-Gordon theory with strongly relevant basic and, at
weak-coupling ($U \ll 2t$), marginally irrelevant double-field
operators.

At $\mu_{\reff}=0$, where the charge excitation spectrum is gapped,
the peculiarity of the charge sector is displayed in the internal
competition of the vacuum configurations of the ordered field driven
by the two sources of gap formation: the ionic term prefers to fix
the charge field at $\sqrt{2\pi K_{c}}\la 0|\varphi_{c}|0\ra
=\pi/2+2\pi n$, which corresponds to the {\it maximum} of the double-field
operators $\sim -U\cos(2\sqrt{2\pi K_{c}}\varphi_{c})$, i.e., it corresponds to a
configuration which is strongly unfavored by the onset of correlations. On the other hand,
the vacuum expectation value of the field $\la 0|\varphi_{c}|0\ra
=2\pi n$, which minimizes the contribution of the double-field operator for $U>0$, leads to the
complete destruction of the CDW pattern, which was favored by the alternating ionic potential.

This type of competition in the double frequency
sine-Gordon model results in  a quantum phase transition from the regime
where the field is pinned in the vacuum of the basic field potential
into the regime where the field is pinned in the vacuum of the
double-frequency cosine term.\cite{Delfino} Qualitatively the
transition point can be estimated from dimensional arguments based
on equating physical masses produced by the two cosine terms. This
allows to distinguish two qualitatively different sectors of the phase
diagram corresponding respectively to the case of weak repulsion
$(U<<\Delta,t)$, where the ground state properties of the system are
determined by the band gap, and to the case of strong repulsion
$(U>>\Delta,t)$, where the charge sector is characterized by the
Mott-Hubbard gap. However, the detailed analysis of the critical area in the case
of the IHM shows, that the BI is separated from the Mott insulator by a
ferroelectric insulating phase.
\begin{figure}[hb]
\vspace{0.2cm}
\includegraphics[width =8.0cm,angle=0]{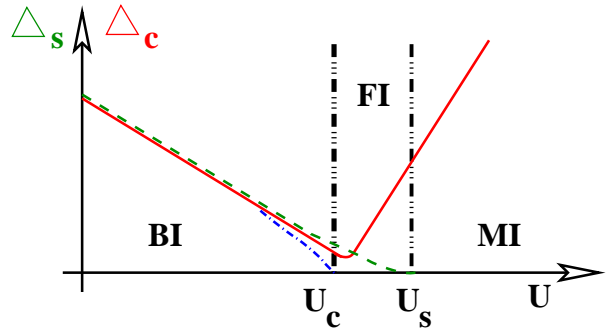}
\caption{Qualitative sketch of behavior of the single particle
(solid line),  spin (dashed line) and optical (dashed-dotted line)
gap as a function of the on-site repulsion $U$  based on the exact
numerical results obtained in Ref. \onlinecite{Brune_03} and Ref.
\onlinecite{Man_04} \label{Fig:Fig2}}
\end{figure}

One important tool to characterize the different phases of the IHM is
to study gaps to excited states, in particular making contact with the gaps obtained in the bosonization description.
Following Ref. \onlinecite{Man_04} we define the optical gap as the
gap to the first excited state in the sector with the same particle
number $N$ and with $S_{z} = 0$, where $S_{z}$ is the $z$-component of the total spin. The single
particle gap is determined as the difference in chemical potential for
adding and subtracting one particle. Finally the spin gap is defined as the energy
difference between the ground state and the lowest lying energy eigenstate in the $S = 1$ subspace.

In Fig. 2 we present a qualitative sketch of the behavior of the single particle charge (solid line),
spin (dashed line) and optical (dashed-dotted line) gap as a function of the on-site repulsion $U$
based on the exact numerical results obtained in
Ref.~\onlinecite{Brune_03} and Ref.~\onlinecite{Man_04}. Two different sectors of the phase diagram
corresponding to the BI and MI are clearly shown. These sectors are distinguished by the pronounced
difference in the $U$ dependence of the charge excitation (single-particle excitation) gap.

Below we treat the ground state phase diagrams of the IHM and of the noninteracting ionic-chain as border
lines of our model. We explore the different character of the excitation gap renormalization by the Hubbard
repulsion $U$ and consider the ground state phase diagram of the half-filled repulsive $t-t^{\prime}$
ionic-Hubbard chain.
The key component of our analysis is based on the assumption that for arbitrary $U$ the infrared properties
of the model are determined by the relation between the controlled effective chemical potential
$\mu_{eff}$ and the value of the charge excitation gap. Moreover, in each charge gapped sector of the phase
diagram, there is only one energy scale which is given either by the ionic distortion or by the Hubbard repulsion.
This allows us to get the qualitative ground state phase diagram, which is summarized in Fig. 3.

At $t^{\prime} < t^{\prime}_{\ast}$ (i.e. $\mu_{\reff}=0$) the
ground state phase diagram remains qualitatively the same as at
$t^{\prime}=0$: the BI phase is separated from a MI phase
via a narrow stripe of the FI phase.

At $\mu_{\reff} \neq 0$, but $t^{\prime} < t^{\prime}_{c}$ the BI phase is
present for weak repulsion. With increasing $U$ the band-gap reduces, which simultaneously leads to a
renormalization of the effective chemical potential $\mu^{r}=2(t^{\prime}-t^{\prime}_{r})$, where
$t^{\prime}_{r}$ is given by (\ref{t^prime_ast}) with $\Delta=\Delta_{r}$. Here we have to consider  two cases
separately.

At $t^{\prime}_{\ast}<t^{\prime}<0.5t$, the renormalized value of the chemical potential always remains less then
the renormalized single-particle gap $\Delta_{r}/2$.
Therefore with increasing $U$ the phase diagram is qualitatively the same as at $t^{\prime} < t^{\prime}_{\ast}$,
i.e.~with increasing $U$ the system undergoes a transition into a FI phase and with further increase of $U$ into a
MI phase. Since the effective single band gap for $t^{\prime}_{\ast}<t^{\prime}$ is smaller
than the single-particle gap at $t^{\prime}=0$, the transition from a BI into the FI insulator takes place for
smaller $U$, which manifests itself in an extension of the FI phase in the direction of lower $U$.

At $0.5t< t^{\prime} < t^{\prime}_{c}$ the BI phase is still realized at weak $U$, but with increasing repulsive
interaction one reaches the critical point $U_{c1}$, where the renormalized value of the chemical potential exceeds
the renormalized single-particle gap $\Delta_{r}/2$. Neglecting quadratic corrections in $\Delta/t$, using
(\ref{IRgaps1}) we easily find that
$
U_{c1} \simeq U^{\ast}[1-8(t^{\prime}-0.5t)/\Delta] \ll U^{\ast}\,.
$
At $U>U_{c1}$ the BI phase is destroyed, the amplitude of the basic frequency field operator
$\sin(\sqrt{2\pi} \varphi_{c})$ in the charge Hamiltonian (\ref{MF_DSG_Charge})
vanishes and the charge sector of the model is given by the Hamiltonian
\bea
{\cal H}_{c} & = &  \int dx
\Big\{\frac{v_{c}}{2}\big[(\partial_{x}\varphi_{c})^2
+ (\partial_x \vartheta_{c})^2 \big] \nonumber \\
&& \hspace{-10mm}  - \mu_{\reff}\sqrt{\frac{2K_{c}}{\pi}}\partial_{x}\varphi_{c}
-\frac{M_{c}}{2 \pi^{2} a_{0}^{2}}\cos(2\sqrt{2\pi
K_{c}}\varphi_{c}) \Big\}\, .
\label{MF_SG_Charge}
\eea
The Hamiltonian  (\ref{MF_SG_Charge}) is a Hamiltonian which describes the charge sector of the
$t-t^{\prime}$ Hubbard chain at $t^{\prime}>0.5t$ and contains two regimes of behavior:\cite{JNB_06}

a) if the effective chemical potential is larger than the value of the correlated (Mott) gap at the transition
point $\mu_{\reff} > M_{c}(U_{c1})$ then the BI phase undergoes a transition into a charge gapless metallic phase.
Since in the BI phase the energy scale of the model is given by the renormalized band-gap, in analogy with the
non-interacting case of ionic chain we expect that the transition from a BI to a metal belongs to the universality
class of commensurate-incommensurate transitions.

b) if $\mu_{\reff} < M_{c}(U_{c1})$ then the BI phase undergoes a transition into a charge gapped ferroelectric phase.

We estimate the charge gap for $U \ll t$  as $M_c  \approx \sqrt{U t}e^{-2\pi t/U}$  and
as  $M_c\approx U$ for $U \gg t$. One finds that for $t^{\prime} \leq t^{\prime}_{c}$ the effective chemical potential
is larger than the exponentially small Hubbard gap $M_{c}(U_{c1})$ and therefore for $0.5t< t^{\prime} < t^{\prime}_{c}$
there opens a window for a transition from the BI to a metallic phase with increasing on-site repulsion. With further
increase of the on-site repulsion, at $U_{c2}$, when $M_{c}(U_{c2}) = \mu_{\reff}$  a charge gap opens once again, and
for $U>U_{c2}$ the system is in the insulating ferroelectric phase (see Fig.3).
\begin{figure}[hb]
\vspace{0.2cm}
\includegraphics[width =8.0cm,angle=0]{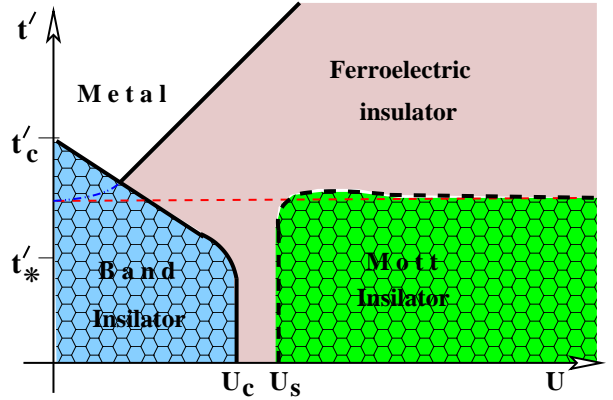}
\caption{Qualitative sketch of the ground state phase diagram of the
$t-t^{\prime}$ ionic-Hubbard chain in the case of repulsive
interaction. Solid lines mark the phase transitions. \label{PD}}
\end{figure}

At $t^{\prime} > t^{\prime}_{c}$ the phase diagram is more simple. Already for $U=0$ the ground state corresponds to a
metallic state, since the effective chemical potential is larger than the band-gap. With increasing Hubbard repulsion a
transition into an insulating phase takes place, when the correlated gap $M_{c}$ becomes larger than
the effective chemical potential. We expect, that similar to the usual $t-t^{\prime}$ model the transition from a metal
to insulator belongs to the universality class of Kosterlitz-Thouless transitions.\cite{AebBaerNoack_01}

Let us now briefly comment on the behavior of the spin sector. Is is usefull first to consider the
strong coupling limit  $U \gg \Delta,t,t^{\prime}$. In this limit the low-energy physics of the $t-t^{\prime}$
ionic-Hubbard chain is described by the spin $S=1/2$ frustrated Heisenberg model
\begin{equation}
 H_{eff}=J\sum_{n}{\bf S}_n\cdot{\bf S}_{n+1}
         +J'\sum_{i}{\bf S}_n\cdot{\bf S}_{n+2}\, ,
\label{SpinHam}
\end{equation}
where the exchange couplings are given by
\be J = \frac{4 t^{2}}{U}\frac{1}{1-\Delta^{2}/U^{2}},\qquad
 J^{\prime}= \frac{4t^{\prime 2}}{U}\, .
 \label{JCouplings}
\ee
For next-nearest neighbor couplings $J^{\prime} <0.25J$ the spin
excitation spectrum of the spin model (\ref{SpinHam}) is gapless and
gapped for $J^{\prime} > 0.25J$.\cite{Haldane} Using
(\ref{JCouplings}) we easily conclude that at $t^{\prime} > 0.5t$
and  $U \gg t,\Delta$ a {\em spin gapped spontaneously dimerized}
phase is realized in the ground state. Since for arbitrary finite
alternating ionic potential the ground state of the system is
characterized by the presence of a long-range ordered CDW
pattern,\cite{Brune_03} we conclude that the whole charge and spin
gapped sector of the phase diagram at $t^{\prime} > 0.5t$
corresponds to a {\em ferroelectric insulating} phase. At
$t^{\prime} < 0.5t$ and for strong repulsion the spin sector is
gapless and therefore in this limit a Mott insulating phase is
realized. Note, that since the ionic potential slightly enhances the
exchange parameter $J$ and does not influence (in first order with
respect to $t^{2}/U$) the next-nearest-neighbor exchange
$J^{\prime}$ for intermediate values of the on-site repulsion $U
\geq 4t$, the Mott phase slightly penetrates into the $t^{\prime} >
0.5t$ sector of the phase diagram.

In the weak-coupling limit at $t^{\prime} > 0.5t$ the FI phase
undergoes a transition either to the metallic phase or directly to
the BI phase. Since the metallic phase with four Fermi points is
characterized by a gapped spin sector for arbitrary weak on-site
repulsion\cite{BalentsFisher_96,Fabrizio_96,DaulNoack_00} we
conclude that the spin gapped phase is a generic feature of the
model for $t^{\prime} > 0.5t$. For $t^{\prime} < 0.5t$, with
increasing $U$ the spin gap continuously decays and finally vanishes
in the MI phase.

To conclude our analysis we briefly discuss the phases which are
realized along the transition lines. The border line between the BI
and metallic phases corresponds to the Luttinger liquid state with
gapless charge and spin excitation spectrum. The border line between
the metallic and FI phases corresponds to the unconventional
metallic phase with gapped spin and completely gapless charge
excitation spectrum. The border line between the BI and FI phases
corresponds to the unconventional metallic phase with gapless
optical excitations and gapped spin and single-particle charge
excitations. Finally, the border line between the FI and MI phases
corresponds to the phase with gapped charge and gapless spin
excitation spectrum.

\section{Conclusions}

We have studied the ground state phase diagram of the half-filled
one-dimensional $t-t^{\prime}$ ionic-Hubbard model using the
continuum-limit bosonization approach. We have shown that the gross
features of the ground state phase diagram and in particular the
behavior of the charge sector  can be described by a quantum
double-frequency sine-Gordon model with topological term. We have
shown that with increasing on-site repulsion, for various values of
the parameter $t^{\prime}$, the model shows the following sequences
of phase transitions: Band insulator -- Ferroelectric Insulator --
Mott Insulator; Band Insulator -- Nonmagnetic Metal -- Ferroelectric
Insulator and Nonmagnetic Metal -- Ferroelectric Insulator.

We expect, that the transition sequence BI-metal-FI found in this
paper is an intrinsic feature not only of the 1D chain, but is a
generic feature of the $t-t^{\prime}$ ionic-Hubbard model and will
show up also in higher spatial dimensions.

\section{Acknowledgments}

It is our pleasure to thank D. Baeriswyl, F. Gebhard, D. Khomskii,
R. Noack, A. Rosch and A.-M. Tremblay for many interesting
discussions. GIJ and EMH acknowledge support from the research
program of the SFB 608 funded by the DFG. GIJ also acknowledges
support from the STCU-grant N 3867.



\begin{thebibliography}{99}
\bibitem{Note_1} A very exceptional case is the 1D electron system with
the short-range (SR) and long-range (LR) Coulomb interaction. As was shown in Ref. \onlinecite{PYMD_97}, in the case of commensurate
band-filling and  SR interaction which support the formation of a
charge-density-wave (CDW) insulating phase, the increasing LR
component of the Coulomb repulsion drives a transition from the CDW
insulator to a metal. Further increase of the LR component leads to
a transition from the metalic phase into the insulating Wigner crystal
phase.
\end{thebibliography}
\end{document}